\documentclass{article}

\usepackage{arxiv}

\usepackage[utf8]{inputenc} 
\usepackage[T1]{fontenc}    
\usepackage[hidelinks]{hyperref}       
\usepackage{url}            
\usepackage{booktabs}       
\usepackage{amsfonts}       
\usepackage{nicefrac}       
\usepackage{microtype}      
\usepackage{graphicx}
\usepackage{natbib}
\usepackage{doi}
\usepackage{subcaption}
\usepackage{amsmath}
\usepackage{authblk}

\title{The effects of turbulence modeling on dynamic stall}

\author[1,*]{Giacomo Baldan}
\author[2]{Francesco Manara}
\author[2,3]{Gregorio Frassoldati}
\author[1]{Alberto Guardone}

\affil[1]{Department of Aerospace Science and Technology, Politecnico di Milano, Milano, Italy 20156}
\affil[2]{Helicopter Division, Leonardo SpA, Samarate, Italy 21017}
\affil[3]{Dipartimento di Ingegneria Civile, Informatica e delle Tecnologie Aeronautiche, Università degli Studi Roma Tre, Roma, Italy 00146}
\affil[*]{Corresponding author: giacomo.baldan@polimi.it}
\date{}

\renewcommand{\headeright}{}
\renewcommand{\undertitle}{}
\renewcommand{\shorttitle}{The effects of turbulence modeling on dynamic stall}

\hypersetup{
pdftitle={The effects of turbulence modeling on dynamic stall},
}

\begin{document}
\maketitle

\begin{abstract}
	A numerical investigation of the flow evolution over a pitching NACA 0012 airfoil incurring in deep dynamic stall phenomena is presented. The experimental data at Reynolds number Re = $1.35 \cdot 10^5$ and reduced frequency $k$ = 0.1, provided by~\citet{Lee2004}, are compared to numerical simulations using different turbulence models. After a preliminary space and time convergence study, two- and three-dimensional URANS with different turbulence models are explored, highlighting the advantages and the drawbacks. Then, the turbulence-resolving capabilities of hybrid RANS/LES strategies are exploited to recover and better represent the dynamic stall vortex. In detail, Scale-Adaptive Simulations (SAS) and Stress-Blended Eddy Simulations (SBES) are adopted. Furthermore, the LES resolved portion allows a spectral analysis of the force and moment coefficients to investigate the contribution of frequency lower than the pitching one. Finally, a comparison of the proposed approaches with other numerical simulations is given.
\end{abstract}

\keywords{Dynamic Stall \and Pitching Airfoil \and Rotorcraft \and Wind Turbine}

\section{Introduction}

Dynamic stall is an unsteady aerodynamic phenomenon arising from rapid changes in the angle of attack of a lifting surface, leading to flow separation.
This phenomenon exhibits complex fluid dynamics phenomena, like shear layers and vortexes, that interact both with each other and with the airfoil~\citep{Smith2020}.
In rotating systems, dynamic stall can be triggered by significant excursions in the angle of attack, blade-vortex interaction, and shocks.
Notably, it becomes particularly sensitive in the context of blades or wings subjected to aeroelastic behavior~\citep{Gardner2019, Hariharan2014}.
The accurate prediction of dynamic stall from the flight envelope is imperative to enhance safety standards and incorporate this phenomenon into the considerations for new designs~\citep{Gardner2023}.
The occurrence of dynamic stall is also one of the critical factor limiting the maximum speed of conventional helicopters~\citep{Baldan2024a, Baldan2024b, Baldan2024c}.

A substantial body of research has been devoted to understanding dynamic stall phenomena through a combination of experimental and computational approaches. 
These investigations have focused on the flow behavior around airfoils undergoing pitching and plunging motions~\citep{Khalifa2021,Zanotti2014a,Avanzi2021,Avanzi2022}.
In recent years, there has been an increase of interest in dynamic stall due to advancements in several key areas.
First, the development of modern GPGPUs has led to significant improvements in computational power and efficiency~\citep{Devanna2022, Devanna2023}.
This, coupled with advances in numerical methods, has enabled more sophisticated simulations of dynamic stall phenomena.
Second, the field has benefited from the development of novel optical measurement methods~\citep{Zanotti2013,Zanotti2014b,Mulleners2013, Gardner2020}.
These techniques in particular provide highly detailed data on the flow around airfoils, which is crucial for validating and improving computational models.
As a result of all these enhancements, researchers have been able to numerically investigate a wide range of factors that influence dynamic stall.
Examples, using wall-resolved LES of ramp-up motion for span-wise-extruded airfoils, include aspect ratio~\citep{Hammer2021, Hammer2022}, sweep angle~\citep{Hammer2023}, and compressibility effects~\citep{Benton2020}.

In this work, we will numerically investigate the experimental campaign conducted over a sinusoidal pitching NACA0012 profile at a Reynolds number of $1.35 \cdot 10^5$ by~\citet{Lee2004}.
Notably, this experimental study has been numerically simulated by several research groups~\citep{Wang2010, Gharali2013, Kim2016, Karbasian2016, Geng2018, Singh2018, Khalifa2023}, but discrepancies are present among all studies.
The differences between numerical predictions and experimental data highlight the need for further investigation and potential refinement of our computational models.
We focus on the influence of turbulence models by employing state-of-the-art formulations.

Section~\ref{sec_numerical_setup} details the computational grid and numerical methods utilized.
Section~\ref{sec_results} presents the results, emphasizing the impact of each model on characteristic dynamic stall phenomena.
Finally, Section~\ref{sec_conclusion} delivers the conclusions drawn from the study.

\section{Numerical setup}\label{sec_numerical_setup}

\begin{figure*}
	\centering
	\includegraphics[width=0.65\textwidth]{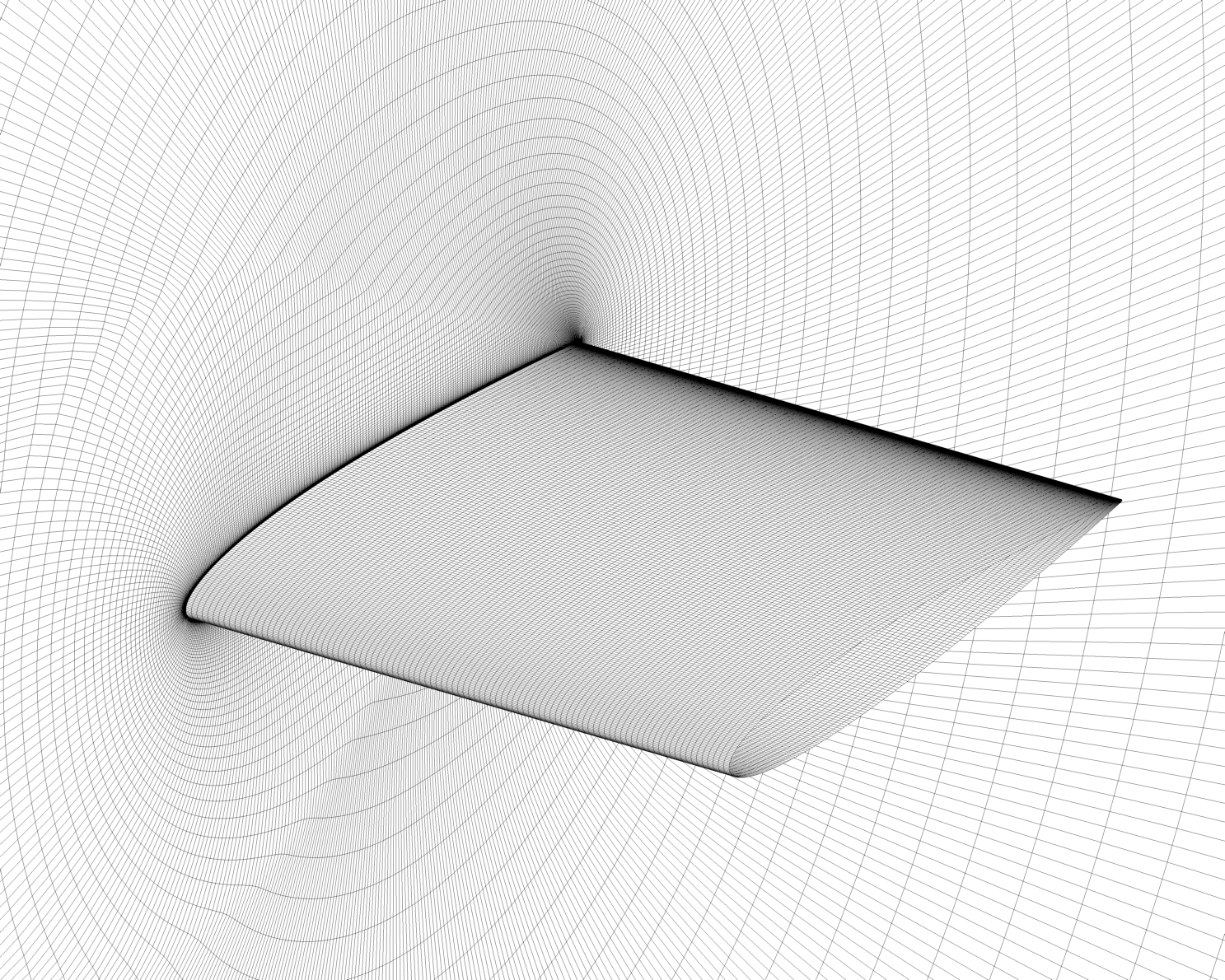}
	
	\caption{Three-dimensional O-grid mesh over the NACA 0012 profile.}
	\label{img_mesh}
\end{figure*}

The numerical setup is configured to match the test conditions in the experimental campaign presented in~\citet{Lee2004}.
A NACA0012 airfoil with 0.15m chord underlying a sinusoidal pitching motion defined as
\begin{align}
	\begin{split}
		\alpha(t) &= \alpha_0 + \alpha_s \sin(\omega t)\\
		&= 10^\circ + 15^\circ \sin \left(18.67 \, [\text{Hz}] \ t\right)
	\end{split}
\end{align}
is investigated, where $\omega$ is the pitching frequency, $\alpha_0$ is the mean angle of attach, and $\alpha_s$ is the angular oscillation amplitude.
The experiments has been carried out, as the present simulations, at Reynolds number Re~=~$1.35\cdot10^5$ and reduced frequency $k=\omega \, c / 2 V_\infty=0.1$.
The free-stream velocity is $V_\infty=14 \, \text{m/s}$ while the pressure is set to $P_\infty=1 \, \text{atm}$.

\subsection{Grid generation}

\begin{table*}
	\caption{Two-dimensional O-grid specifications.}
	\label{table_grid}
	\begin{center}
		\begin{tabular}{*{8}{c}}
			\toprule
			Name & $N_\xi$ & $N_\eta$ & Nodes & Quads & $\Delta \xi_{le} \ (\cdot \, 10^{-3}  \, \text{c})$ & $\Delta \xi_{te} \ (\cdot \, 10^{-4}  \, \text{c})$ & Growth rate \\
			\midrule
			$\mathcal{G}1$ & 384 & 96 & \ \,36\,768 & \ \,36\,290 & 3.33 & 8.45 & 1.20\\
			$\mathcal{G}2$ & 512 & 128 & \ \,65\,408 & \ \,64\,770 & 2.00 & 5.00 & 1.10\\
			$\mathcal{G}3$ & 1024 & 256 & 261\,888 & 260\,610 & 0.67 & 3.67 & 1.05\\
			\bottomrule
		\end{tabular}
	\end{center}
\end{table*}

A preliminary two-dimensional space convergence study has been performed using three different O-grid meshes.
Structured meshes have been adopted because they grant a higher solution accuracy compared to unstructured ones.
Indeed, to reach the same solution quality an unstructured-polyhedral grid requires from two to six times the number of hexahedra in a delayed detached eddy simulation (DDES)~\citep{Kozelkov2016}.
O-grids have been chosen in this work because the highly stretched elements in the wake, especially near the trailing edge, present in C-grid resulted in non-physical patterns in the flow fields.
The strong anisotropy in the wake region influences the angle at which the trailing edge stall occurs resulting in mispredicted loads.
An expanding fan with origin at the sharp trailing edge has been implemented to mitigate the artificial flow patterns though they are still noticeable.
Even considering the mean pitching angle as the direction where the wake is more present is not sufficient to solve the issue.
A last reason to choose O-grid is that usually C-grids are adopted to better catch the flow behavior in the wake, thus having a more accurate prediction of the drag.
However, when considering dynamic stall, the most significant contribution to the loads is due to the pressure drop of the dynamic stall vortex and the high angle of attack.

The O-grids are obtained through a hyperbolic extrusion in the normal direction starting from the profile surface.
The wall-normal spacing follows a geometric progression with growth rates that ranges from 1.2 for the coarse grid $\mathcal{G}1$ to 1.05 for the fine grid $\mathcal{G}3$.
Another critical aspect in the space convergence, especially for dynamic stall simulations, is the element size at the leading and at the trailing edges.
The leading edge spacing is extremely important since it covers the region where the laminar separation bubble resides during the pitch up phase and subsequently the dynamic stall vortex is generated.
The same consideration can be applied also to the trailing edge since stall can also start from the back of the profile.
Also for this reason, in all meshes, the blunt trailing edge is jointed with a circle that avoids the imposition of the separation point and at the same time increases the orthogonality of the grid allowing a faster convergence.
The $\eta^+ < 1$ requirement is always satisfied at the wall, even for the most demanding flow condition when the acceleration, due to the high angle of attack, increases the velocity at more than twice the free-stream value. Thus, the first layer is positioned at $5.0\cdot10^{-6} \ \text{m}$ over the surface.  Table~\ref{table_grid} reports the details of the grids, where $\xi$ is the coordinate over the profile and $\eta$ is the direction normal to the surface.

The three-dimensional grid is obtained through a uniform extrusion of the $\mathcal{G}2$ grid in the span-wise direction.
According to the best practice, presented in~\citet{Shur1999}, the span-wise size is equal to one chord to fully describe the dynamic stall vortex.
The number of points, instead, is retrieved respecting the LES requirements~\citep{Choi2012}, and is set equal to 150.
The extruded three-dimensional mesh has a total number of 9.9M points and 9.73M hexahedra.
Figure~\ref{img_mesh} reports a graphical representation.

\subsection{Numerical methods}
\begin{figure}
	\centering
	\includegraphics[width=0.5\linewidth]{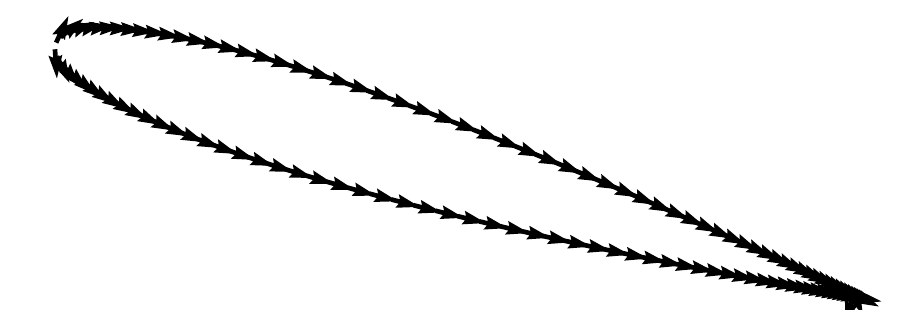}\\
	\caption{Example of the curvilinear coordinate used in the $C_f$ computation. The origin is located at the leading edge while the end is positioned in the middle of the trailing edge for both sides of the airfoil.}
	\label{img_geometric_direction}
\end{figure}

\begin{figure}
	\centering
	\includegraphics[width=0.5\linewidth]{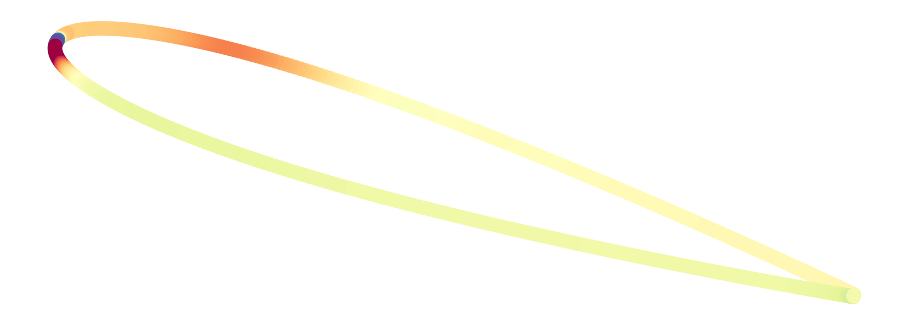}\\
	\vspace*{4mm}
	\includegraphics[width=0.4\linewidth]{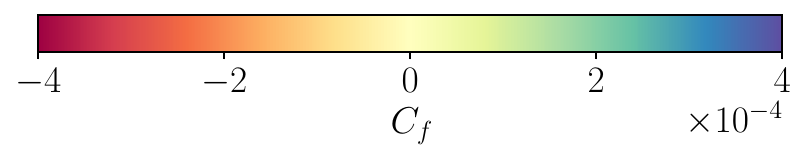}\\
	\caption{Example of the post-processed $C_f$ distribution. It is noticeable the discontinuity at the leading edge due to the origin of the coordinate system and the recirculation portion on the suction side of the profile due to the backward advection of the dynamic stall vortex.}
	\label{img_cf_example}
\end{figure}

The steady and unsteady incompressible RANS equations are solved using \textit{ANSYS Fluent 2023R2}~\citep{ANSYS}.
A second-order upwind discretization in space is adopted.
Gradients are retrieved through a least square cell-based method, and fluxes are obtained with the Rhie-Chow momentum-based formulation.
The SIMPLE method is leveraged to solve the pressure-velocity equations.
The time evolution is obtained through a second-order implicit time integration scheme.

Moving on to boundary conditions, the farfield marker present in the original mesh is split in two parts.
For this reason, a marker is set for cells with $x>0$, corresponding to the coordinate around which the grid rotates and being $x$ the direction of the free-stream flow.
Farfield faces are thus separated using the mark imposing a velocity inlet and a pressure outlet boundary conditions.
The inflow velocity is imposed together with the turbulent properties that correspond to a turbulence intensity of 0.08\%, as in the wind tunnel, a turbulent to laminar viscosity ratio of 0.1 and an intermittency level equal to 0.1.
A non-slip and non penetrating condition is imposed on the airfoil. 
When the three-dimensional mesh is considered the same boundary conditions apply in the extruded surfaces, while a translational conformal periodic condition is imposed at the planar extremities of the domain to grant a nominally infinite wing.

RANS system is closed using three different turbulence models.
As a first guess, a standard $k \omega$-SST turbulence model~\citep{Menter1994} has been tested with the low-Reynolds correction.
Second, the previous model has been investigated with the addition of the $\gamma$ intermittency equation~\citep{Menter2004}, in which a differential transport equation is solved for the intermittency, to better describe the transition of the boundary layer from laminar to turbulent.
Finally, the Reynolds Stress Model~(RSM) of \citet{Launder1975} has been analyzed as an alternative to the previous ones, based on the Boussinesq hypothesis, to improve the description of the separated region that plays a major role in this type of flow.

Moving to three-dimensional hybrid simulation, Scale Adaptive Simulations~(SAS)~\citep{Menter2010} and Stress Blended Eddy Simulations~(SBES)~\citep{Menter2018} are leveraged to recover the scale resolving capabilities in the Large Eddy Simulation region.
A critical aspect of these methods is that they rely on standard RANS turbulence methods in the portion near the profile where the grid requirements otherwise would be unfeasible for most engineering applications.
This implies that the boundary layer transition still relies on RANS turbulence modeling making the choice of the model a key aspect.

The airfoil pitching motion is prescribed through the rigid motion of the entire grid.
Indeed, the angular velocity is imposed in \textit{Fluent} using a user-defined expression equal to $\dot \alpha(t) = \omega \ \alpha_s \cos(\omega t)$.
The mean rotation, instead, is granted through a rigid rotation of the mesh before starting the simulation.
In order to speed up temporal convergence between subsequent cycles, the simulation is performed in three different stages. 
Firstly, a steady RANS simulation with the mesh rotated at $\alpha_0$ is performed until force coefficient convergence is reached. 
Note that the solution at this stage does not affect particularly the unsteady solution.
After that, the time-dependent pitching simulation is evolved.
The mesh rotation is imposed to obtain a negative pitch rate after the steady simulation~\citep{Baldan2023}.
This allows to reach unsteady convergence faster since the flow is evolved at lower angle of attack and is attached to the profile.
Usually, a quarter of a cycle and the entire following one are neglected to reach convergence.
Finally, an entire pitching cycle is computed for URANS simulations while multiple cycles are performed for hybrid RANS/LES simulations.

Following each simulation, the pressure coefficient $C_p$ and skin friction coefficient $C_f$ distributions over the airfoil profile are extracted for each time step within a single pitching cycle.
These unstructured outputs from \textit{Fluent} are subsequently mapped onto the structured mesh of the airfoil using a k-d tree algorithm, which efficiently identifies the nearest neighbors~\cite{Virtanen2020}.
For post-processing, the skin friction value at each point is converted to its signed magnitude.
The sign is determined based on the direction calculated from the $x$ and $y$ components of $C_f$.
The positive geometric direction aligns with the curvilinear coordinate that extends from the leading edge to the trailing edge, as illustrated in Figure~\ref{img_geometric_direction}.
When the geometric direction and the numerically computed $C_f$ direction coincide, the sign is positive. 
Conversely, if they oppose each other, the sign is negative. 
This process allows for the identification of regions with flow separation.
Figure~\ref{img_cf_example} reports an example in which the recirculation portion can be seen on the suction side of the profile due to the backward advection of the dynamic stall vortex.

\section{Results}\label{sec_results}
The present section discusses the results obtained with the previously described numerical model. 
In detail, the space and time convergence study (Section~\ref{ssec_convergence}), the turbulence models comparison for pure RANS in two dimensions (Section~\ref{ssec_2d_rans}), the differences using three-dimensional simulations and hybrid RANS/LES techniques (Section~\ref{ssec_hybrid_results}), analysis using the SBES model of the cycle-to-cycle variations including the spectral analysis of the loads (Section~\ref{ssec_sbes}), and the comparison of the proposed computations with other numerical simulations present in literature (Section~\ref{ssec_other_simulations}).

\subsection{Space and time convergence study}\label{ssec_convergence}

\begin{figure*}
	\centering
	\subfloat{\includegraphics[width=0.48\textwidth]{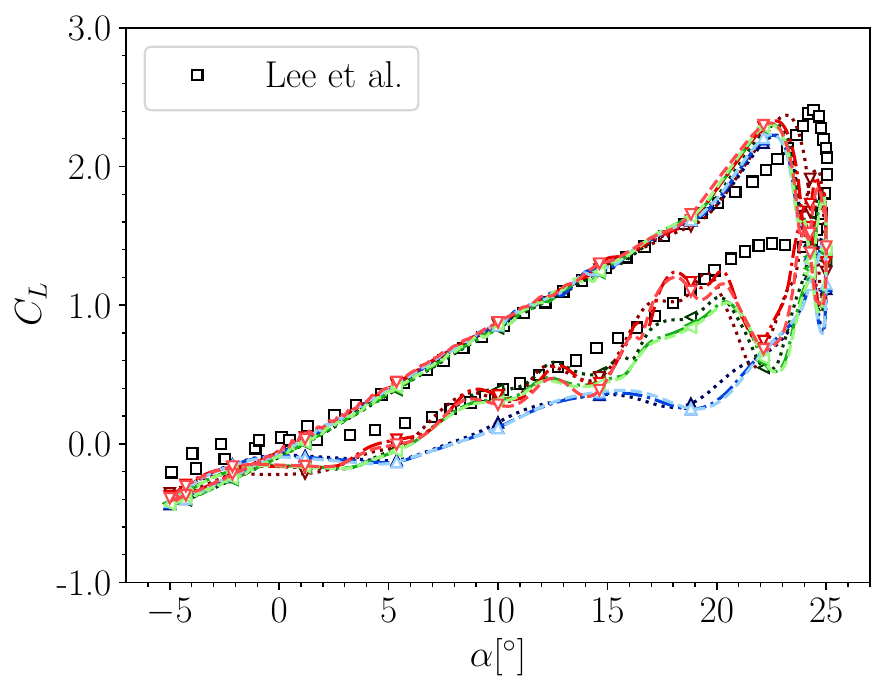}}
	\hfill
	\subfloat{\includegraphics[width=0.48\textwidth]{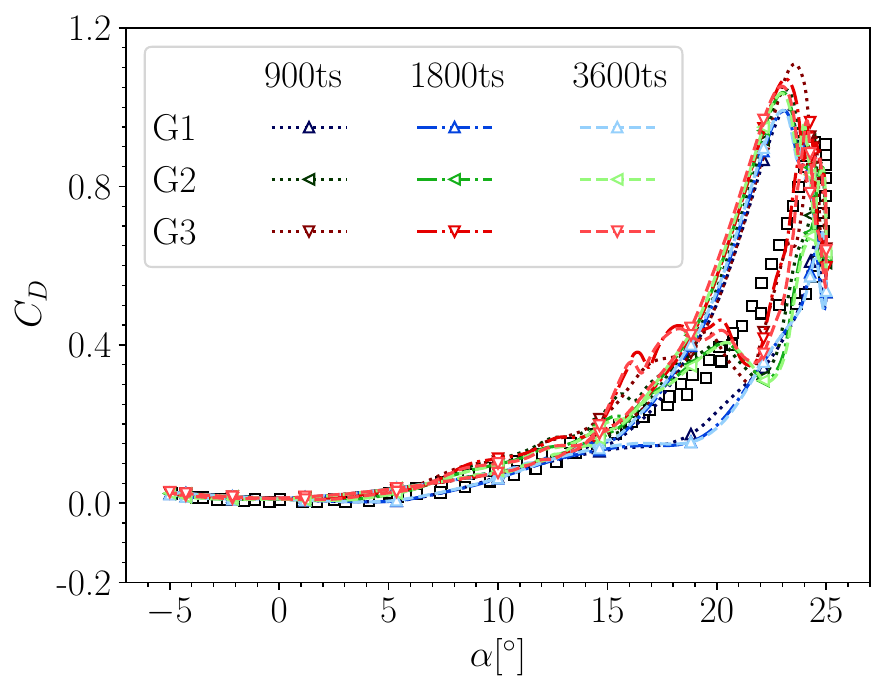}}\\
	
	\caption{Convergence study of lift and drag coefficients using $\mathcal{G}1$, $\mathcal{G}2$, and $\mathcal{G}3$ grids and 900, 1800, and 3600 time steps per cycle. Experimental data from~\citet{Lee2004}.}
	\label{img_convergence_2d}
\end{figure*}

A comprehensive grid and time step sensitivity analysis was conducted to ensure the numerical accuracy of the simulations.
Three computational meshes, namely $\mathcal{G}1$, $\mathcal{G}2$, and $\mathcal{G}3$, with varying resolutions were implemented, discretized with 900, 1800, and 3600 time steps per pitching cycle, consisting of a total of nine simulations.
The SST turbulence model with the intermittency equation was employed to close the RANS equations.

The solution convergence is evaluated looking at the lift and drag profile over the entire pitching cycle.
Figure~\ref{img_convergence_2d} represents the analyzed loads.
Firstly, we focused on the upstroke region and at the pitching angle at which lift and drag peaks corresponding to the generation of the dynamic stall vortex (DSV).
All three grids are almost overlapping in the up-stroke phase showing scarce sensitivity to space and time resolution.
However, when we look at the vortex evolution and convection over the airfoil grid spacing plays a major role. 
On the contrary, the time step size has only a limited impact on the predicted loads.
It should be noted that the solver reaches a faster convergence when using a larger number of time steps per cycle.
Grid $\mathcal{G}1$ is deemed inadequate due to its insufficient resolution, resulting in a notable discrepancy in the down-stroke behavior. 
This difference is attributed to the inability of the coarse grid to accurately capture the evolution of the DSV.
For these reasons, the following computations are performed using the medium grid, thus $\mathcal{G}2$, and 3600 time steps per cycle.

\subsection{Two-dimensional URANS}\label{ssec_2d_rans}

\begin{figure*}
	\centering
	\subfloat{\includegraphics[width=0.48\textwidth]{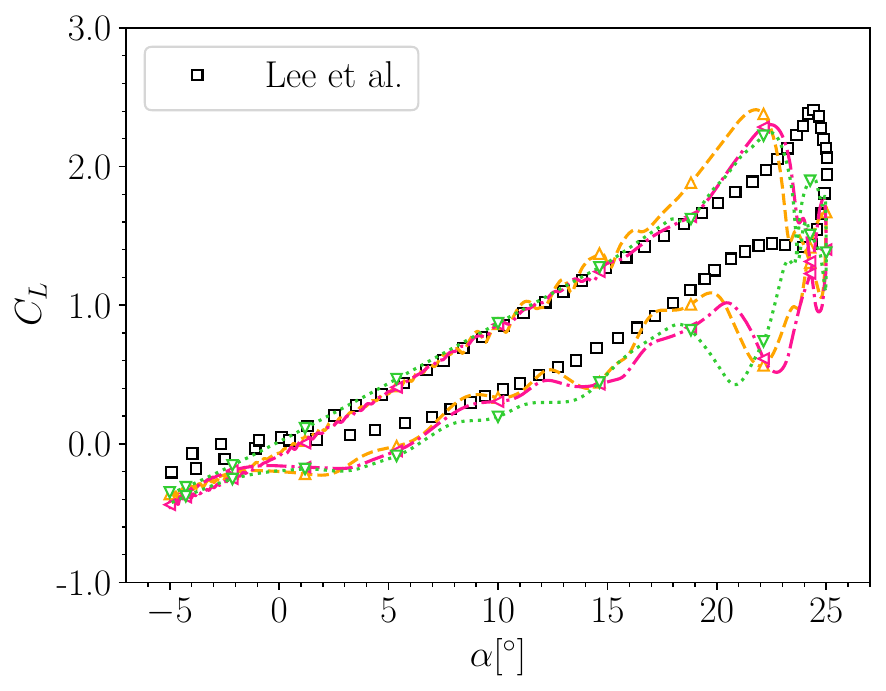}}
	\hfill
	\subfloat{\includegraphics[width=0.48\textwidth]{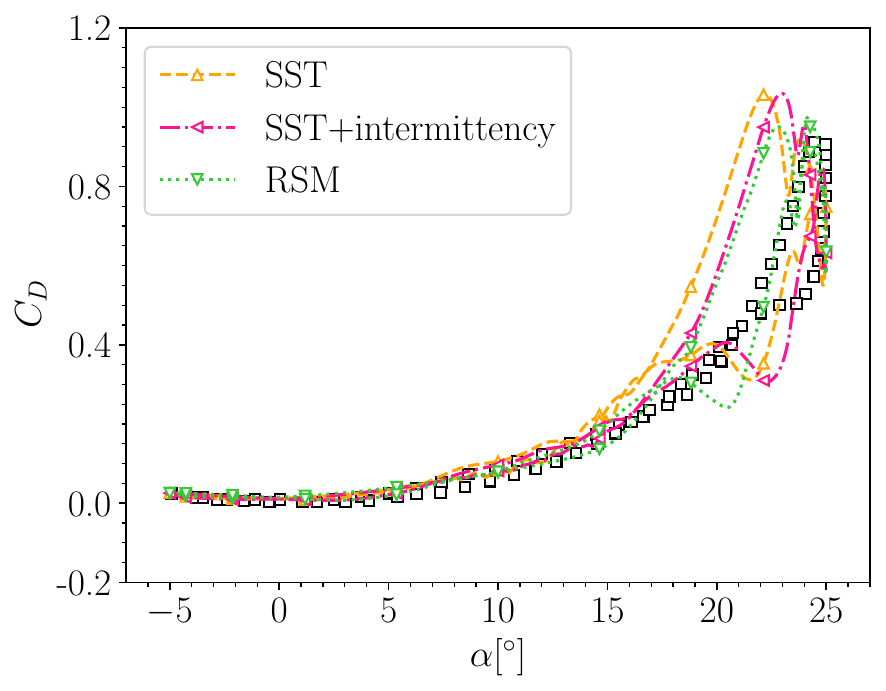}}\\
	
	\caption{Two-dimensional turbulence model comparison of lift and drag coefficients using $\mathcal{G}2$ grid and 3600 time steps per cycle. A standard SST turbulence model~\citep{Menter1994} is compared against SST+intermittency~\citep{Menter2004} and Reynolds Stress Model~(RSM)~\citep{Launder1975}. Experimental data from~\citet{Lee2004}.}
	\label{img_model_comp_2d}
\end{figure*}

\begin{figure*}
	\centering
	\subfloat[SST]{\includegraphics[width=0.33\textwidth]{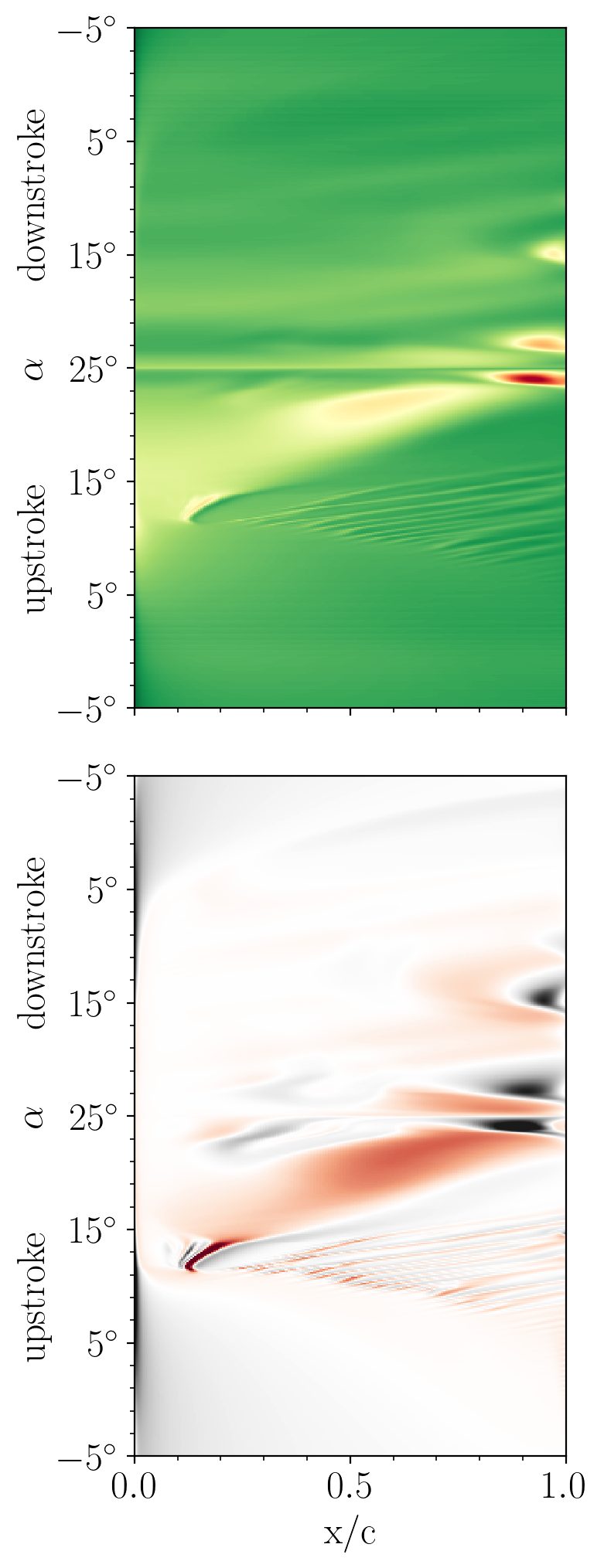}}
	\hfill
	\subfloat[SST+intermittency]{\includegraphics[width=0.33\textwidth]{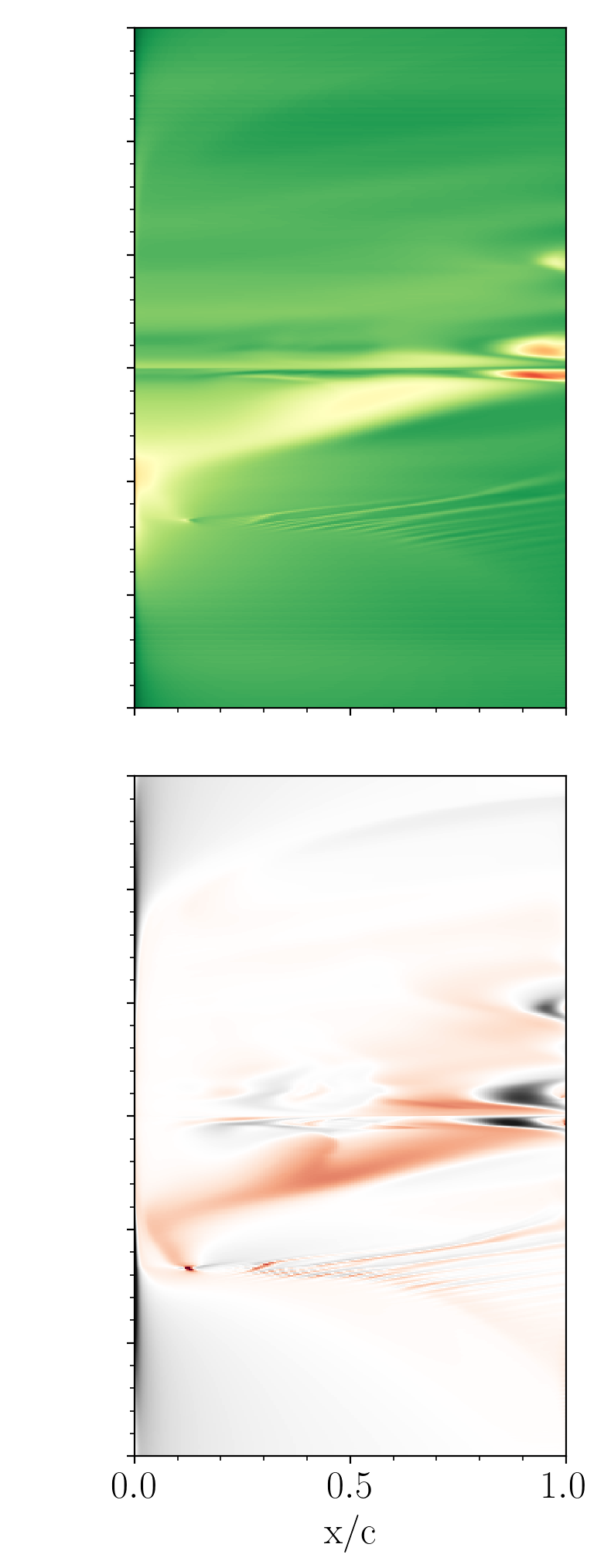}}
	\hfill
	\subfloat[RSM]{\includegraphics[width=0.33\textwidth]{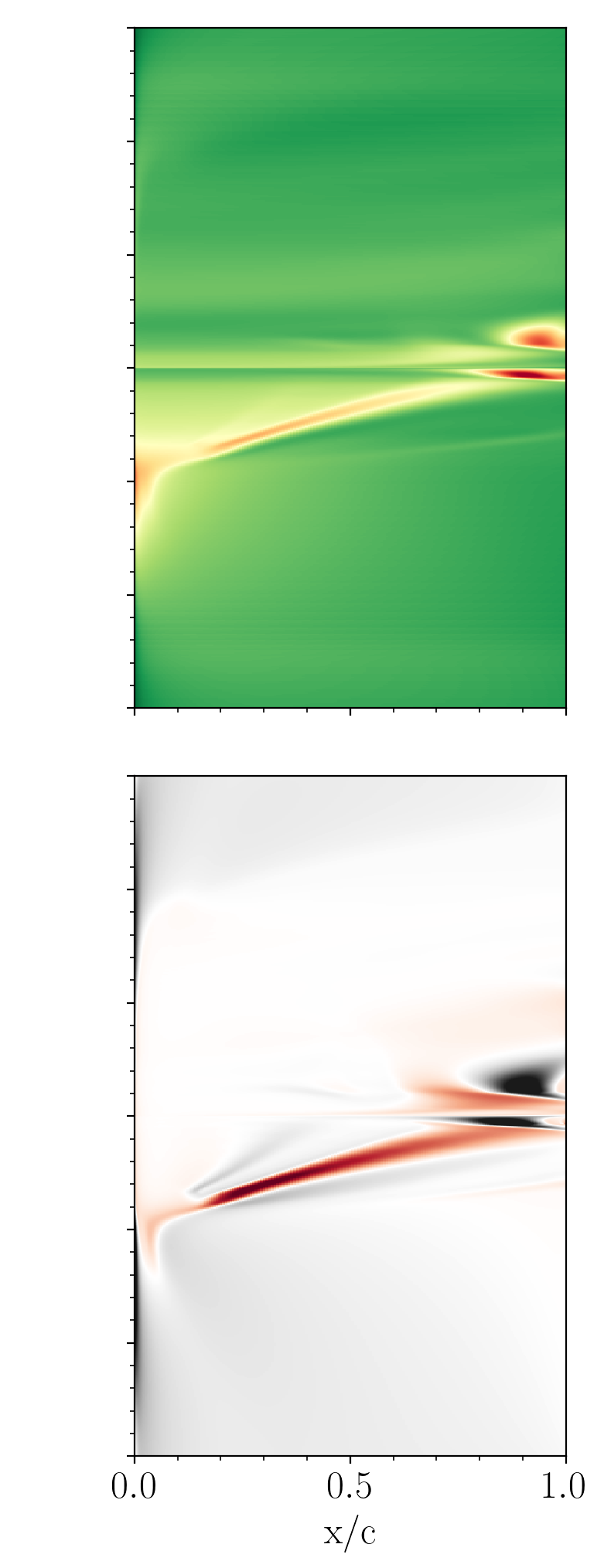}}\\
	\vspace*{5mm}
	\subfloat{\includegraphics[width=0.4\textwidth]{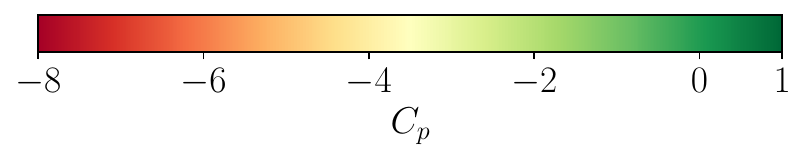}}
	\hspace*{1cm}
	\subfloat{\includegraphics[width=0.4\textwidth]{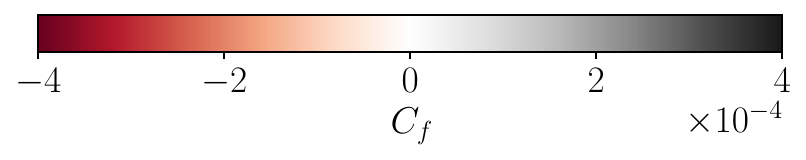}}
	
	\caption{Pressure and friction coefficient over the suction side of the airfoil for the two-dimensional  test cases.}
	\label{img_2d_imshow}
\end{figure*}

\begin{figure*}
	\centering
	\subfloat[SST]{\includegraphics[width=\textwidth]{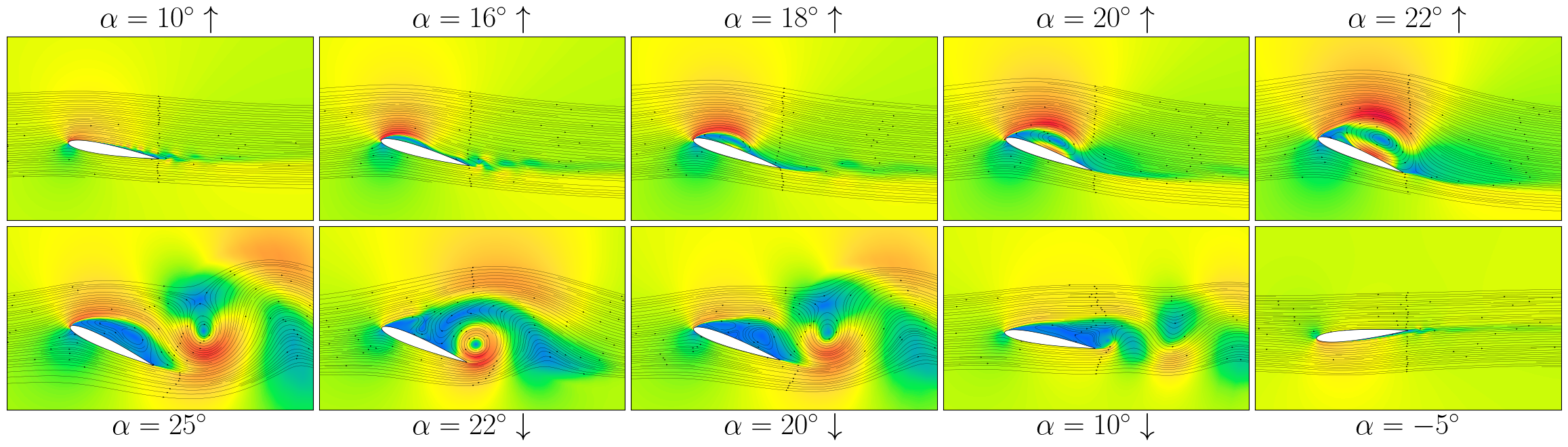}}\\
	\vspace*{5mm}
	\subfloat[SST+intermittency]{\includegraphics[width=\textwidth]{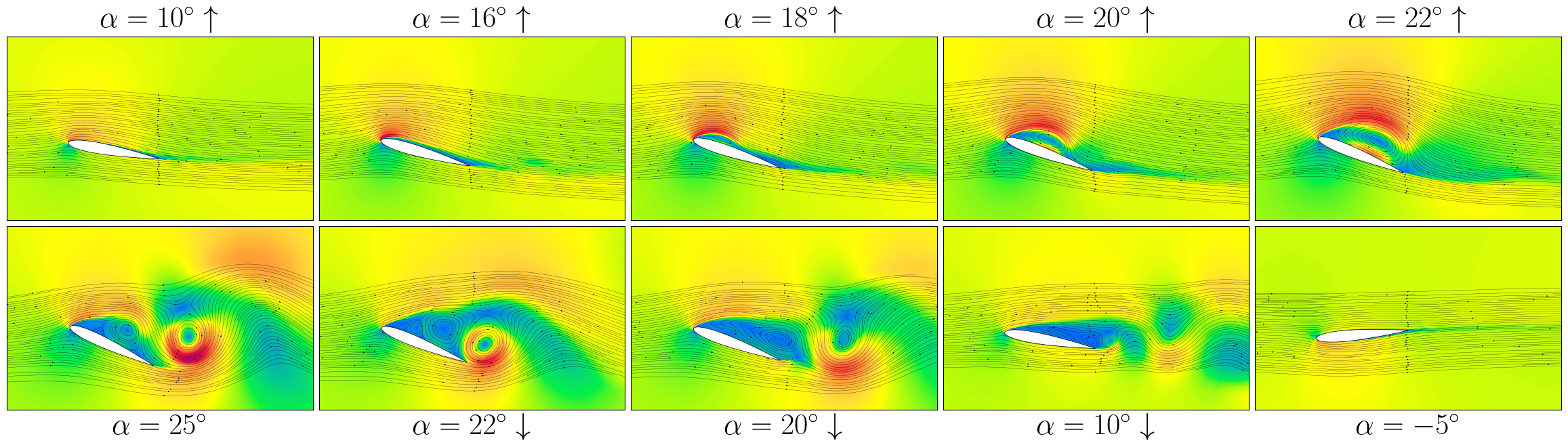}}\\
	\vspace*{5mm}
	\subfloat[RSM]{\includegraphics[width=\textwidth]{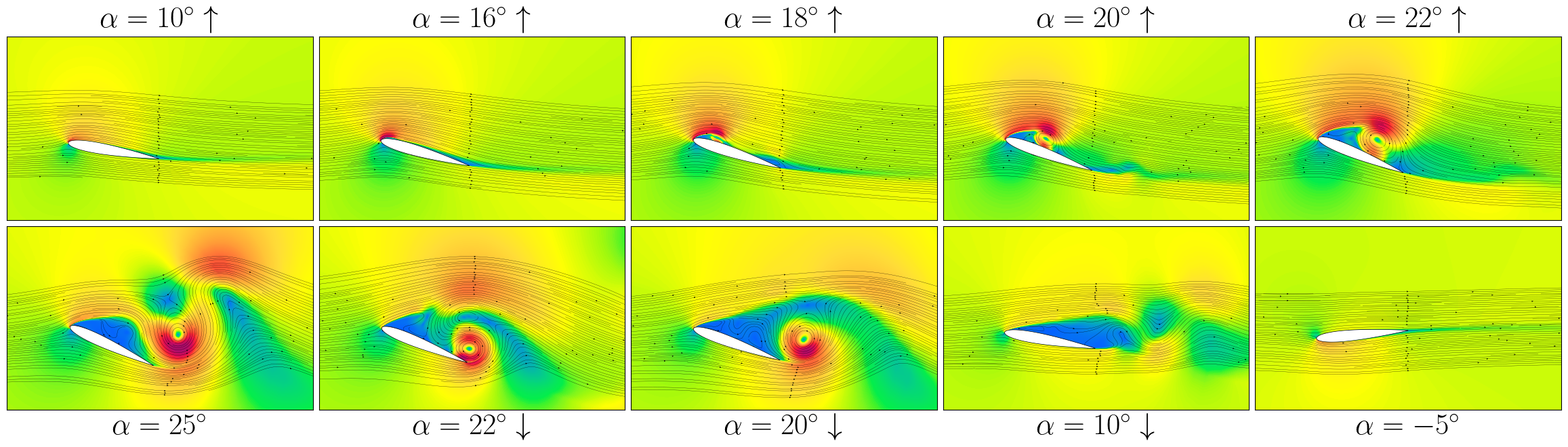}}\\
	\vspace*{5mm}
	\subfloat{\includegraphics[width=0.4\textwidth]{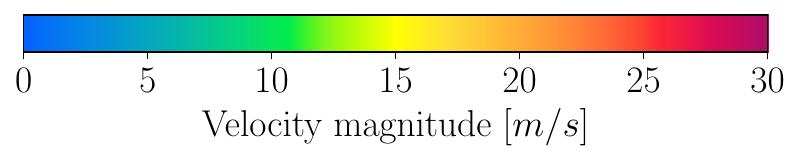}}
	
	\caption{Instantaneous velocity magnitude contour and streamlines for the two-dimensional test cases.}
	\label{img_model_comp_velocity_contour_2d_isst}
\end{figure*}

A two-dimensional URANS approach is used in the initial study of the turbulence model effects. 
The investigation focuses on three models: the standard $k \omega$-SST with low-Re correction~\citep{Menter1994}, the same model with an additional $\gamma$ intermittency equation solved through a transport equation~\citep{Menter2004}, and the Reynolds Stress Model (RSM)~\citep{Launder1975}.
This analysis concentrates on three key aspects: the evolution of airfoil loads as a function of angle of incidence, the pressure coefficient $C_p$ and skin friction coefficient $C_f$ distributions over the suction side of the profile, and the velocity magnitude contours for the entire flow field at characteristic angles of dynamic stall.

The first aspect we focus on is the effect on lift and drag.
Figure~\ref{img_model_comp_2d} presents the evolution of the forces as a function of the angle of attack. Notably, during the upward pitching phase, the standard SST model exhibits oscillatory behavior beyond 10 degrees.
Additionally, it predicts the earlier development of the dynamic stall vortex compared to the other two models, leading to significantly higher peak values for both $C_L$ and $C_D$ coefficients.
Conversely, the SST model with intermittency and the Reynolds Stress Model showcase remarkably similar results, with the RSM predicting marginally lower peak drag.
However, a crucial discrepancy arises when comparing all three models against experimental data by~\citet{Lee2004}.
All models predict DSV formation at a lower angle of attack, leading to a completely stalled region at the top of the airfoil, which was not observed in the experiments.
During the downward pitching phase, all models exhibit comparable behavior and demonstrate reasonable agreement with the experimental data, excluding the initial portion characterized by high angles of attack, where the flow remains completely stalled.

Turning our attention to Figure~\ref{img_2d_imshow}, which illustrates pressure and skin friction coefficients on the profile's suction side, we observe pronounced differences between the various turbulence models.
A distinct contrast lies in the absence of instabilities within the Reynolds Stress Model (RSM), unlike the two SST models.
Furthermore, the depiction of the laminar separation bubble varies significantly.
In the standard SST model, at angles of attack between 11-14 degrees, a substantial recirculation zone arises near the leading edge and expands markedly in size.
Incorporating transition into the SST model alters this behavior, the recirculation portion becomes more localized and coincides with instabilities originating from the trailing edge at approximately 12 degrees.
Finally, the RSM model demonstrates notably lower pressure coefficient $C_p$ values, resulting from its lack of interaction with the instabilities.
Furthermore, a significant distinction lies in the contrasting behaviors observed in the vortex backward propagation and the subsequent flow separations at the trailing edge.
For the initial two turbulence models, the region characterized by negative skin friction coefficient encompasses a substantial area, impacting a large portion of the upper surface. 
Additionally, these models predict three distinct detachments originating from the trailing edge, exhibiting progressively decreasing intensity as the incidence of the profile diminishes.
In contrast, the RSM model portrays a more localized effect of the DSV evolution and results in only two detachments from the trailing edge.
Finally, an examination of the velocity magnitude contours in Figure~\ref{img_model_comp_velocity_contour_2d_isst} corroborates several aspects emphasized in the analysis of previous figures. Notably, the fluctuations in profile loads around 10-degree angle of attack, observed in the standard SST model, are visually evident within the profile's wake. Furthermore, the discrepancy in the DSV size is apparent between the SST and RSM models; in the latter, the LSB detaches from the profile as early as 18 degrees of incidence. 

In general, all three analyzed models successfully capture the essential elements of dynamic stall, LSB formation and subsequent DSV generation, however they exhibit limitations in representing certain aspects of the flow evolution. The standard SST model demonstrates shortcomings in accurately predicting boundary layer transition as well as the RSM model. On the contrary, the SST model with intermittency equation offers improved LSB description. Nonetheless, when significant flow detachment occurs, the SST models struggle to represent the flow evolution accurately.

\subsection{Three-dimensional hybrid RANS/LES}\label{ssec_hybrid_results}

\begin{figure*}
	\centering
	\subfloat{\includegraphics[width=0.48\textwidth]{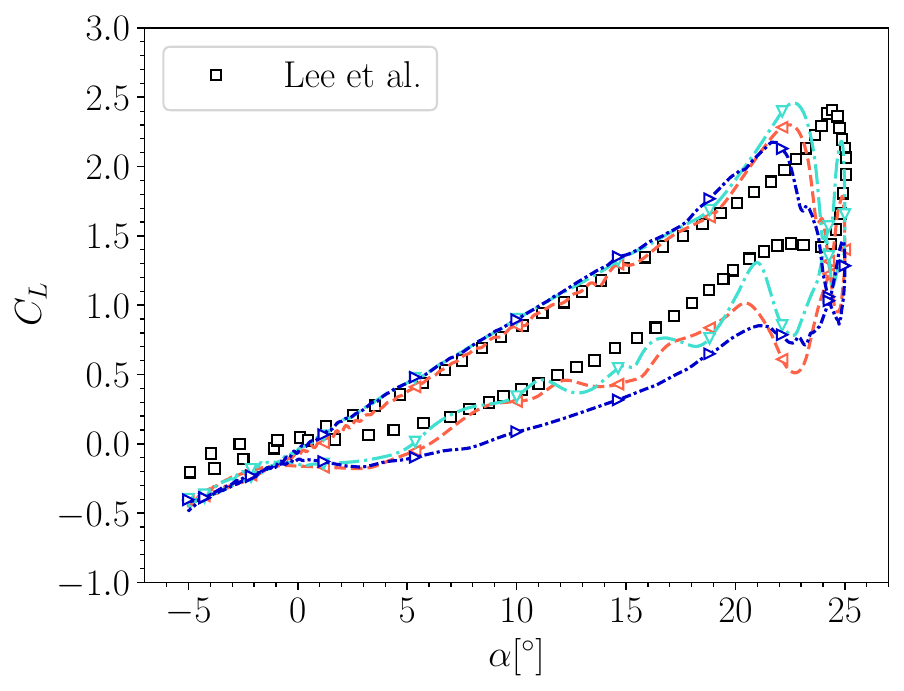}}
	\hfill
	\subfloat{\includegraphics[width=0.48\textwidth]{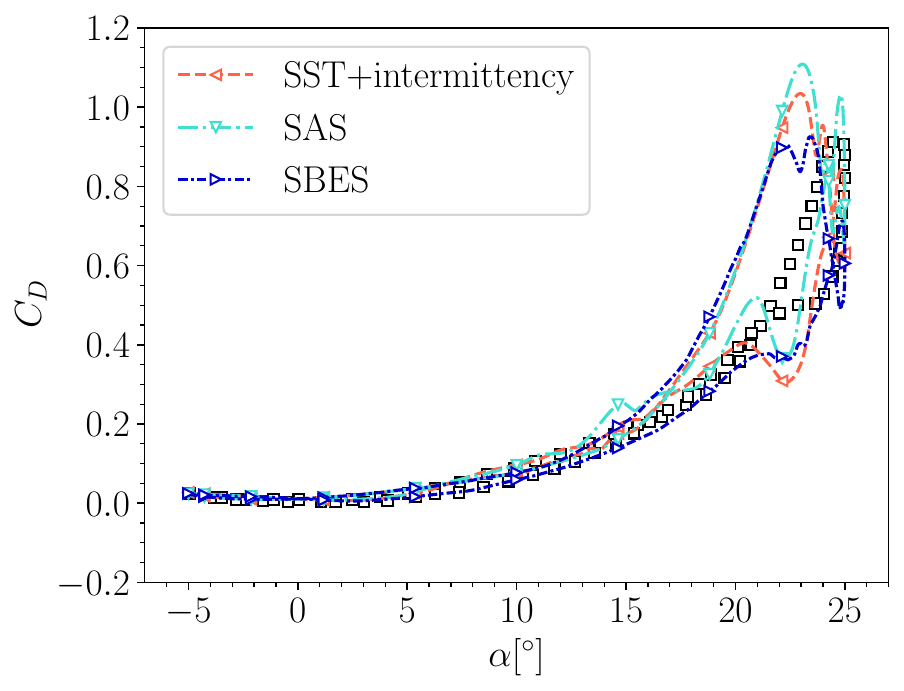}}\\
	
	\caption{Three-dimensional turbulence model comparison of lift and drag coefficients. A pure RANS approach with SST+intermittency~\citep{Menter2004} turbulence model is compared with two hybrid RANS-LES techniques: Scale Adaptive Simulations~(SAS)~\citep{Menter2010} and Stress Blended Eddy Simulations~(SBES)~\citep{Menter2018} turbulence model. Experimental data from~\citet{Lee2004}.}
	\label{img_model_comp_3d}
\end{figure*}

\begin{figure*}
	\centering
	\subfloat[SST+intermittency]{\includegraphics[width=0.33\textwidth]{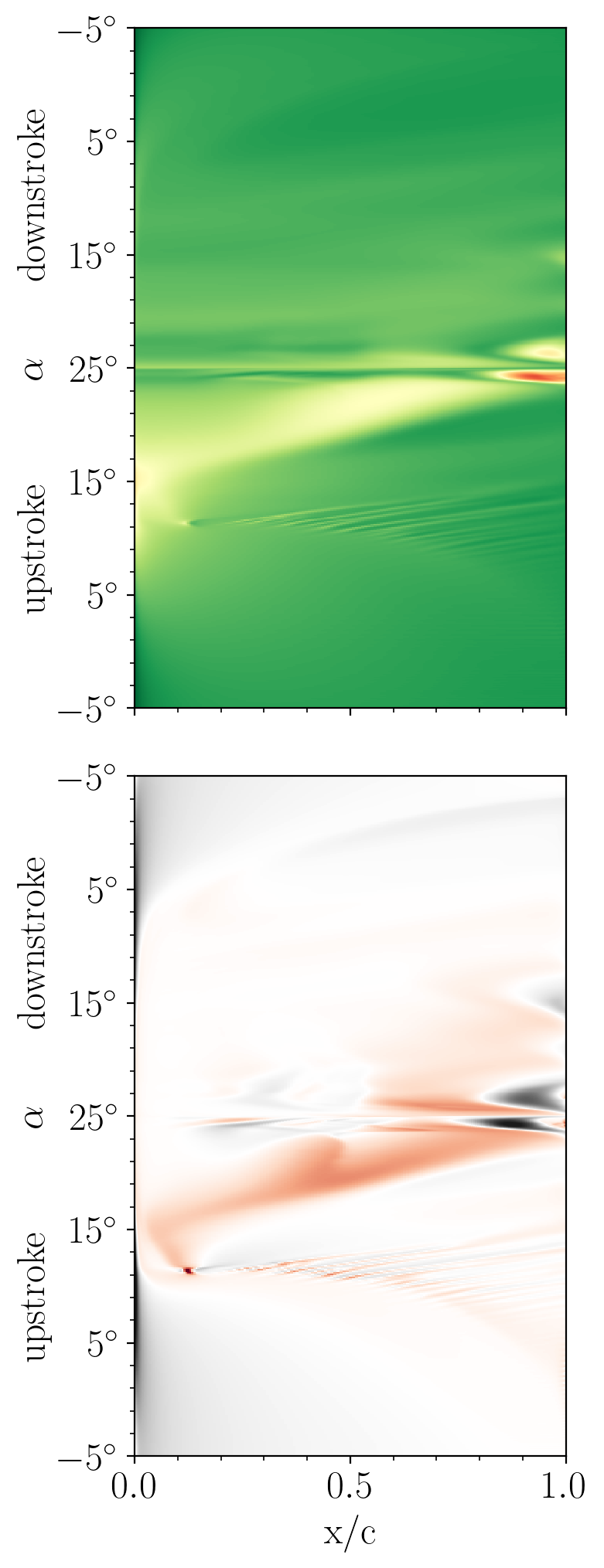}}
	\hfill
	\subfloat[SAS]{\includegraphics[width=0.33\textwidth]{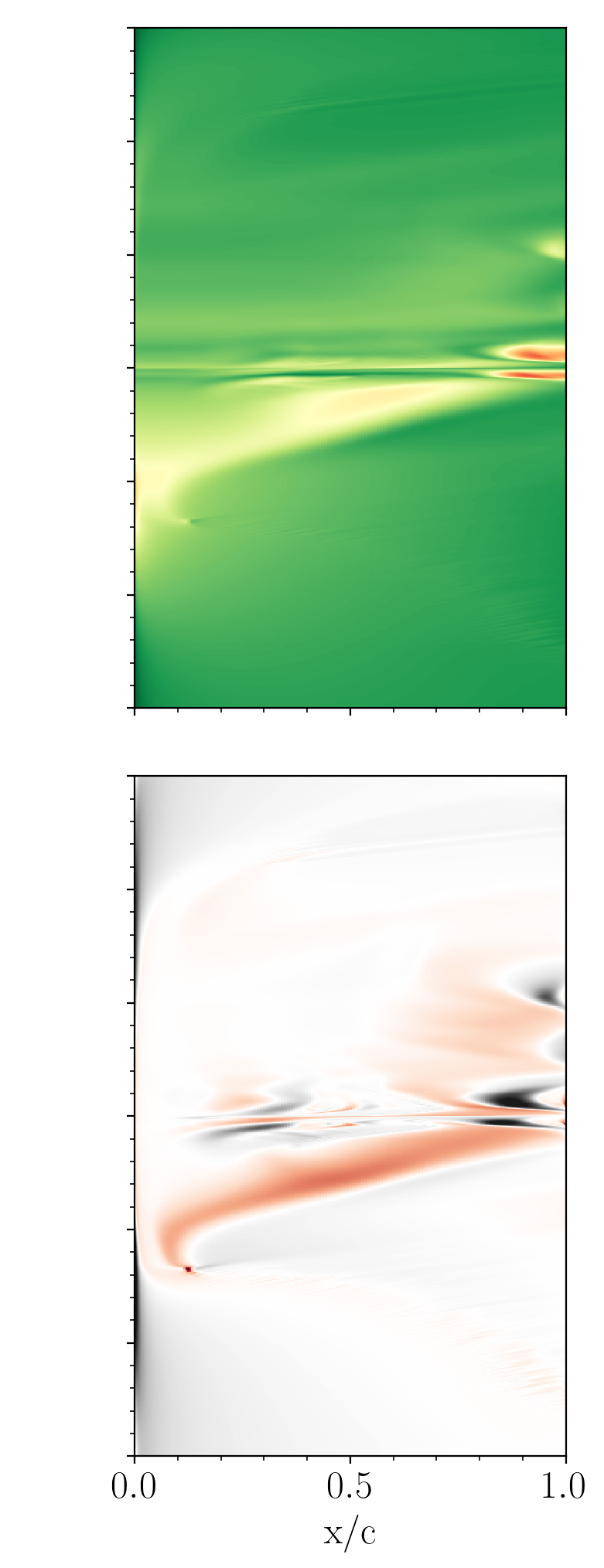}}
	\hfill
	\subfloat[SBES]{\includegraphics[width=0.33\textwidth]{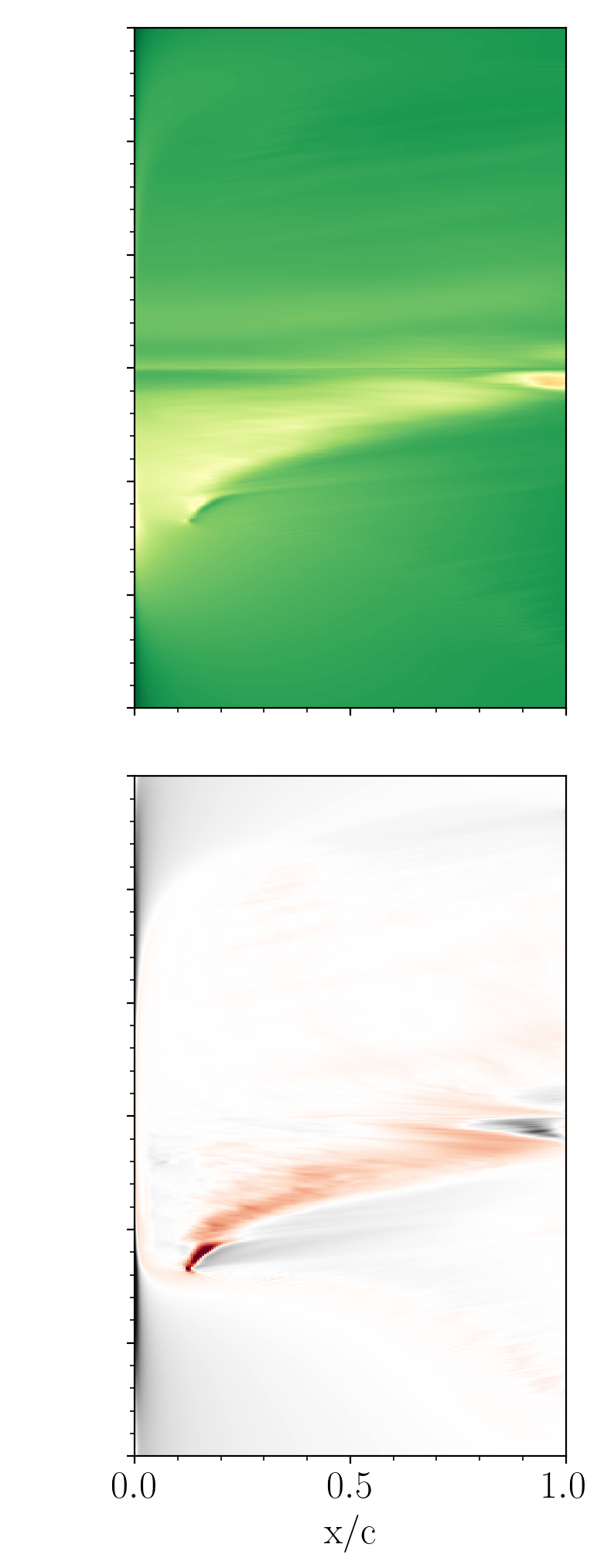}}\\
	\vspace*{5mm}
	\subfloat{\includegraphics[width=0.4\textwidth]{images/cmap_imshow_cp.pdf}}
	\hspace*{1cm}
	\subfloat{\includegraphics[width=0.4\textwidth]{images/cmap_imshow_cf.pdf}}
	
	\caption{Pressure and friction coefficient over the suction side of the airfoil for the three-dimensional test cases.}
	\label{img_3d_imshow}
\end{figure*}

\begin{figure*}
	\centering
	\includegraphics[width=\textwidth]{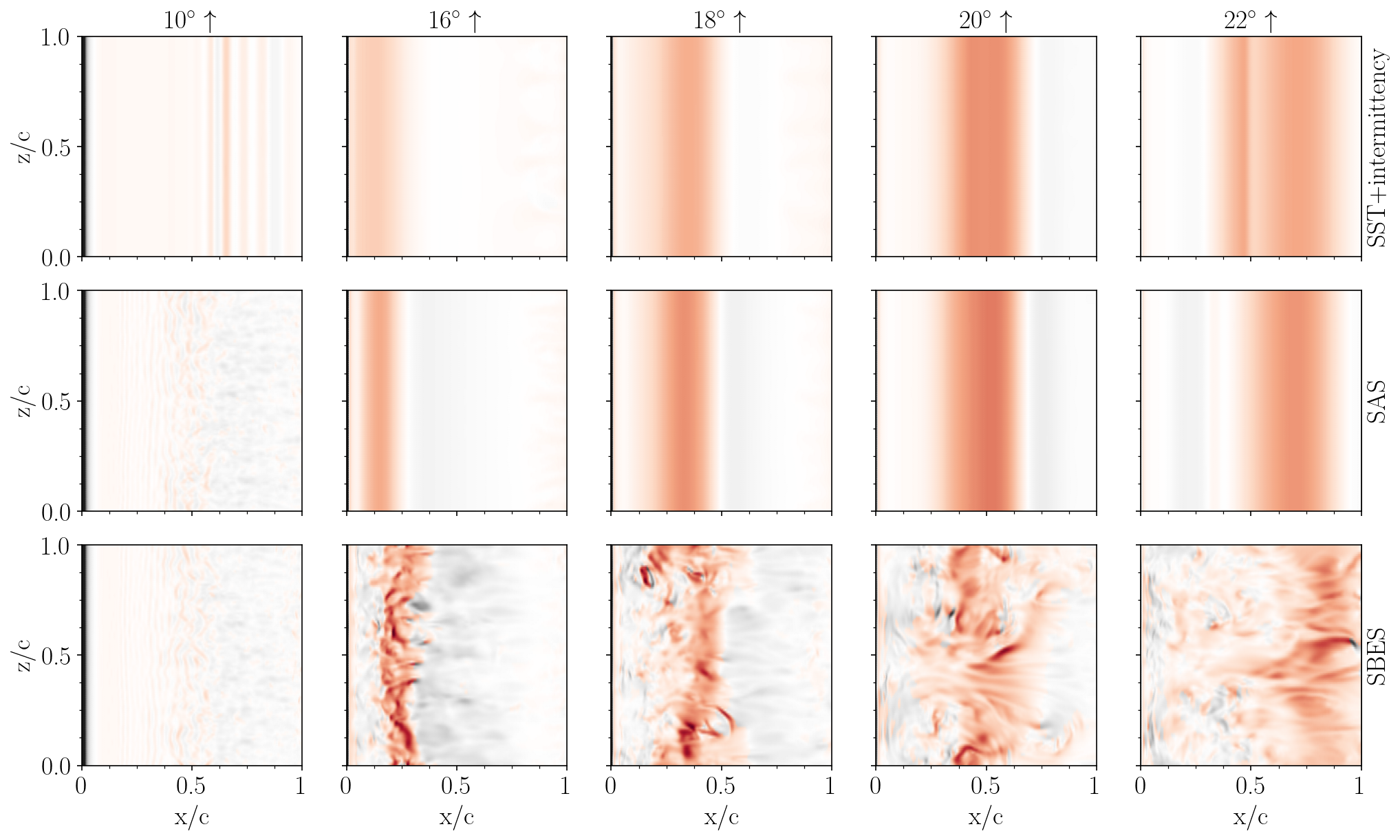}\\
	\vspace*{6mm}
	\includegraphics[width=\textwidth]{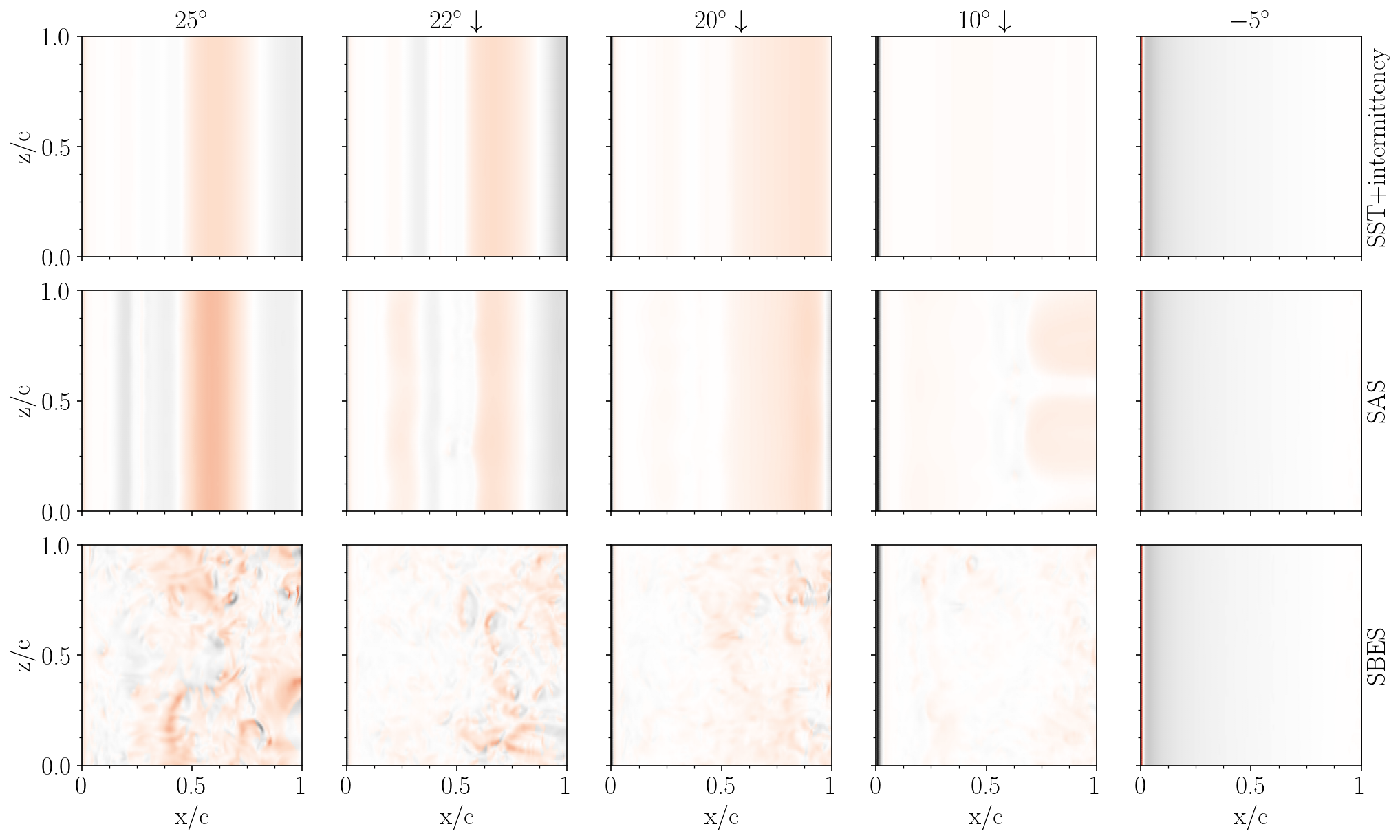}\\
	\vspace*{1mm}
	\includegraphics[width=0.4\textwidth]{images/cmap_imshow_cf.pdf}
	
	\caption{Instantaneous skin friction coefficient contours on the suction side of the extruded airfoil for the three-dimensional test cases ($z$ is the span-wise direction).}
	\label{img_3d_cp_cf}
\end{figure*}

\begin{figure*}
	\centering
	\subfloat[SST+intermittency]{\includegraphics[width=\textwidth]{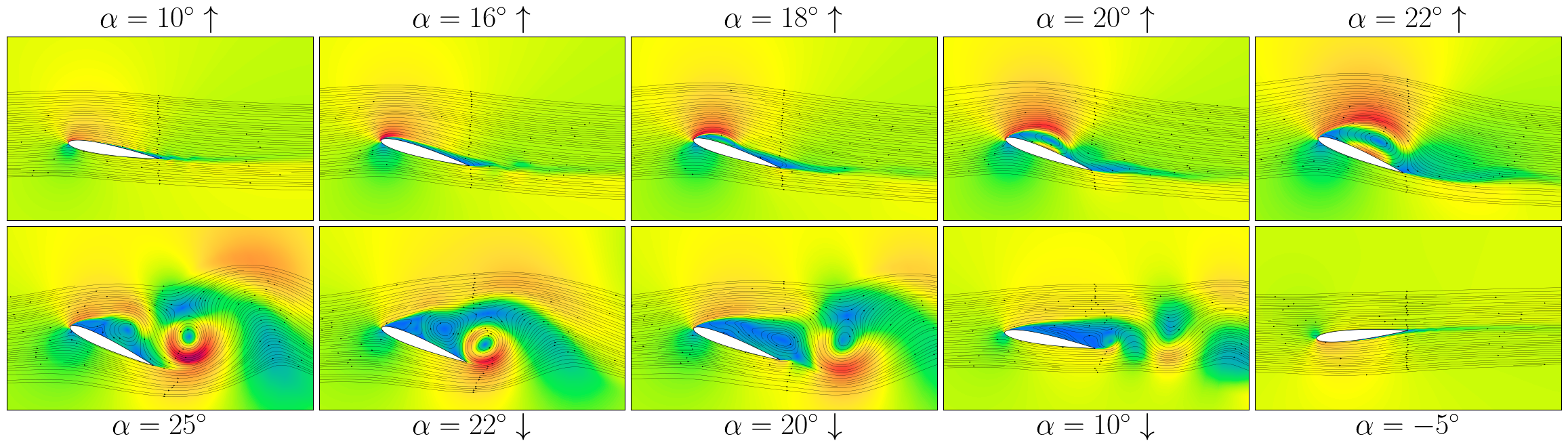}}\\
	\vspace*{5mm}
	\subfloat[SAS]{\includegraphics[width=\textwidth]{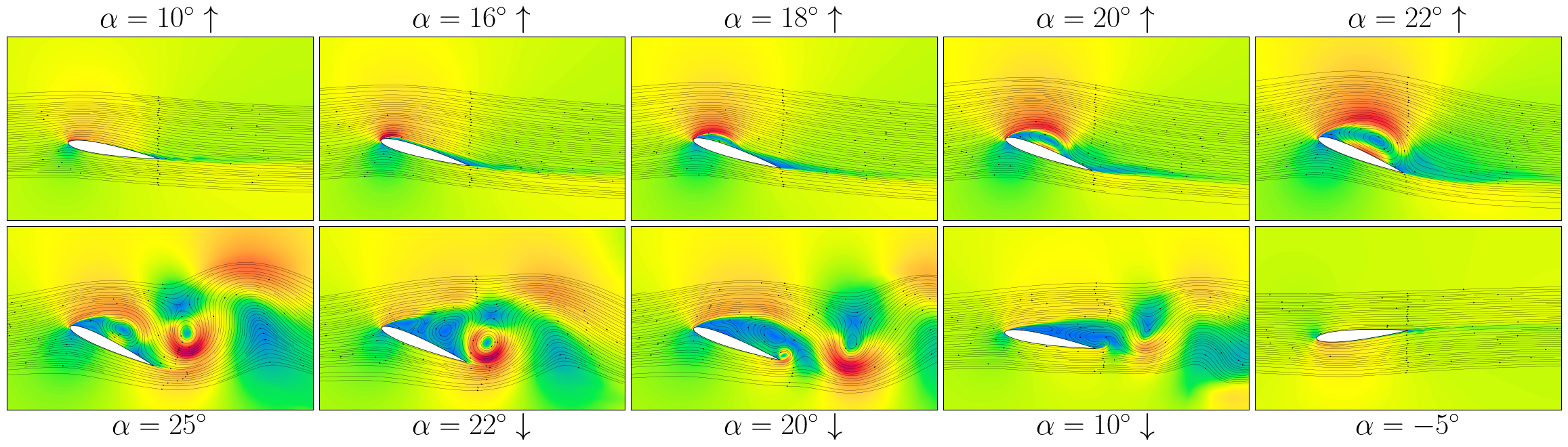}}\\
	\vspace*{5mm}
	\subfloat[SBES]{\includegraphics[width=\textwidth]{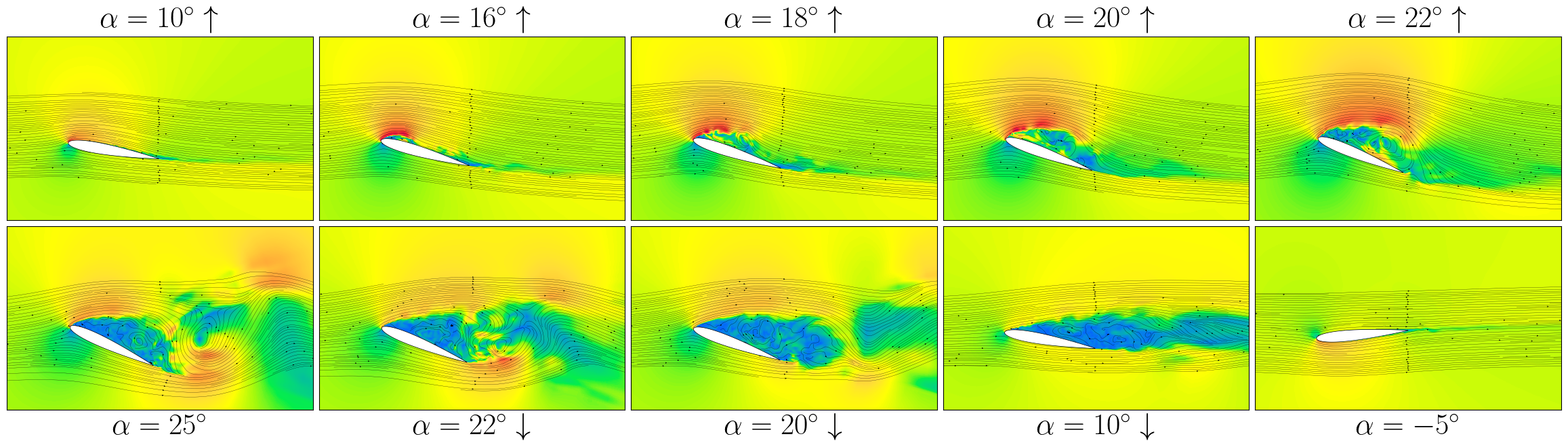}}\\
	\vspace*{5mm}
	\subfloat{\includegraphics[width=0.4\textwidth]{images/model_comp_velocity_contour.pdf}}
	
	\caption{Instantaneous velocity magnitude contour and streamlines for the three-dimensional test cases.}
	\label{img_model_comp_velocity_contour_3d}
\end{figure*}

Considering the inherently three-dimensional nature of dynamic stall~\citep{Ullah2024}, and of turbulence phenomena in general, a more rigorous analysis was carried out through the use of three-dimensional simulations. 
As an initial comparison, a URANS simulation utilizing the SST model with the intermittency equation with the low-Re correction was included as in the two-dimensional test case.
To achieve a more accurate description of the stalled region and the dynamic stall vortex, hybrid RANS/LES techniques were leveraged. 
Specifically, the Scale Adaptive Simulations~(SAS)~\citep{Menter2010} and the Stress Blended Eddy Simulations~(SBES)~\citep{Menter2018}  models were utilized.
The RANS portion of the hybrid model leveraged the same turbulence model as the pure RANS simulation, which, based on the obtained results, offered the most accurate representation of the boundary layer transition.
Beyond the three key aspects employed in conventional two-dimensional simulations, this study incorporates additional investigations.
One such analysis focuses on the instantaneous distribution of the skin friction coefficient distribution on the suction side of the extruded profile.
This evaluation aims to highlight the impact of the transition function used when shifting from a RANS model to a LES model on the solution behavior in the span-wise direction.

Figure~\ref{img_model_comp_3d} displays the lift and drag coefficients obtained for the first converged pitching cycle.
Notably, for hybrid methods, the solution exhibits cycle-to-cycle variations, particularly during the downward pitching phase.
While attached flow conditions lead to consistent results across all methods, deviations arise upon LSB detachment and subsequent DSV formation. 
The SAS model overestimates the forces, while the SBES model underestimates them, compared to the pure SST model.
This aligns with the previously observed angle misprediction in two-dimensional simulations, where the backward convection of the DSV deviates from experimental data of~\citet{Lee2004}.
Conversely, analyzing the portion where the flow is completely stalled with negative pitching velocity, the SBES model displays a trend closer to the experimental data, albeit with a persistent offset.
In contrast, the other two models, while closer to the experiments in terms of values, exhibit significant oscillations attributable to their representation of the DSV.

Figure~\ref{img_3d_imshow} provides insights into the load behavior by depicting the $C_p$ and $C_f$ distributions on the upper surface of the extruded wing. 
The RANS and SAS simulations exhibit similar trends, characterized by three distinct vortexes shedding from the trailing edge. 
Additionally, the upstroke region and LSB separation show close agreement between these models. 
A closer examination reveals that the hybrid model predicts fewer oscillations associated with the laminar-to-turbulent boundary layer transition. 
A noticeable difference arises near the maximum angle of attack, where the hybrid model predicts higher pressure peaks. 
The SBES model, on the other hand, exhibits significantly different behavior, as previously observed in the load analysis. 
The first key observation is the absence of trailing-edge vortex shedding, with the exception of the primary DSV originating from the nose. 
Furthermore, the DSV description is fundamentally different, with turbulent flow fluctuations captured in the $C_f$ distribution.

Analyzing the span-wise distribution of $C_f$ on the wing's upper surface, Figure~\ref{img_3d_cp_cf}, through instantaneous contours further validates the preceding observations.
The SST model incorporating intermittency predicts no fluctuations along the span-wise direction ($z$-axis), essentially reducing the solution to a purely two-dimensional case. 
The SAS model, instead, exhibits minimal span-wise $C_f$ fluctuations when the DSV is present, as observed at angles of attack of $10^\circ$ and $22^\circ$ during the downstroke phase. 
In contrast, at $\alpha=10^\circ$ during the upstroke phase, it exhibits a pattern characteristic of a LES approach. 
Finally, the SBES model demonstrates superior turbulence resolving capabilities throughout most of the flow field. 
During the descent phase, the flow exhibits the physical characteristics of a fully developed stall.
The skin friction distribution reveals that the SAS-predicted vortex shows a higher magnitude compared to the other models. 
Additionally, the SBES model displays a peak located rearwards compared to the others.
Notably, the SBES model possesses a crucial advantage: the ability to capture the entire physics of the DSV. 
During the upstroke phase at $\alpha=20^\circ$ and $22^\circ$, the span-wise extent of the vortex appears comparable to the wing profile chord length, indicating a more realistic representation of the flow phenomenon.

Examining the instantaneous velocity fields presented in Figure~\ref{img_model_comp_velocity_contour_3d} reveals that the SBES model predicts a lower maximum velocity compared to the other two models.
This observation aligns with the previously discussed characteristic of the SBES model, where it captures a larger portion of the resolved flow structures.

\subsection{Cycle-to-cycle variation analysis}\label{ssec_sbes}

\begin{figure*}
	\centering
	\includegraphics[width=\textwidth]{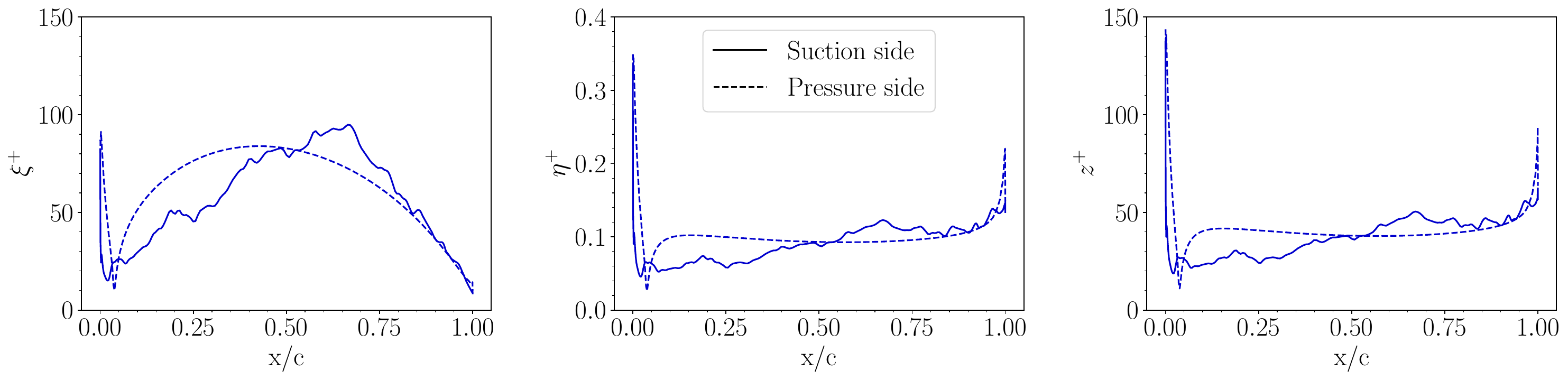}
	
	\caption{Span-wise averaged plus units at maximum incidence for SBES test case.}
	\label{img_sbes_plus_unit}
\end{figure*}

\begin{figure*}
	\centering
	\subfloat{\includegraphics[width=0.48\textwidth]{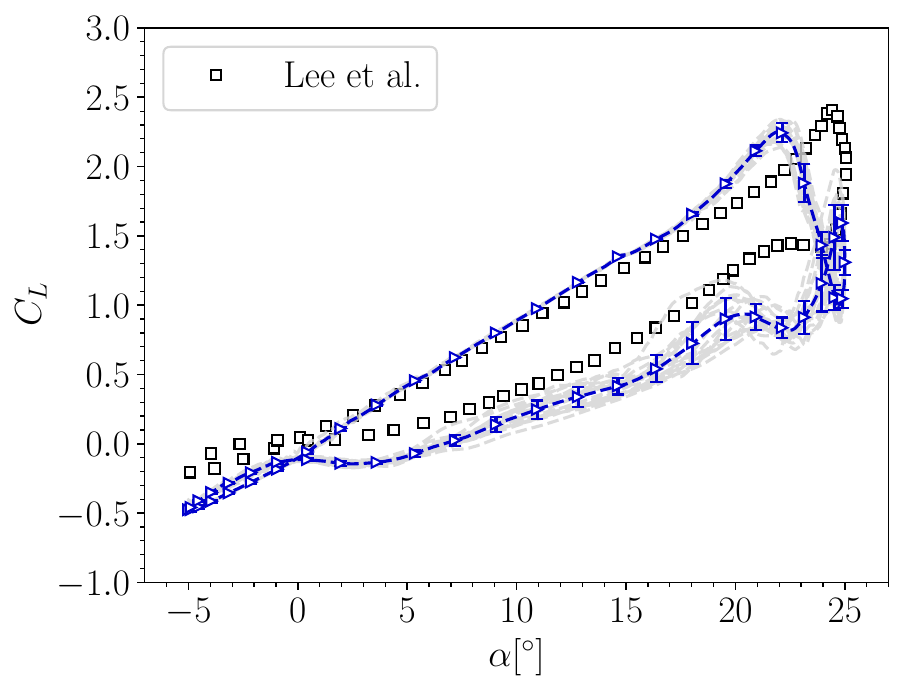}}
	\hfill
	\subfloat{\includegraphics[width=0.48\textwidth]{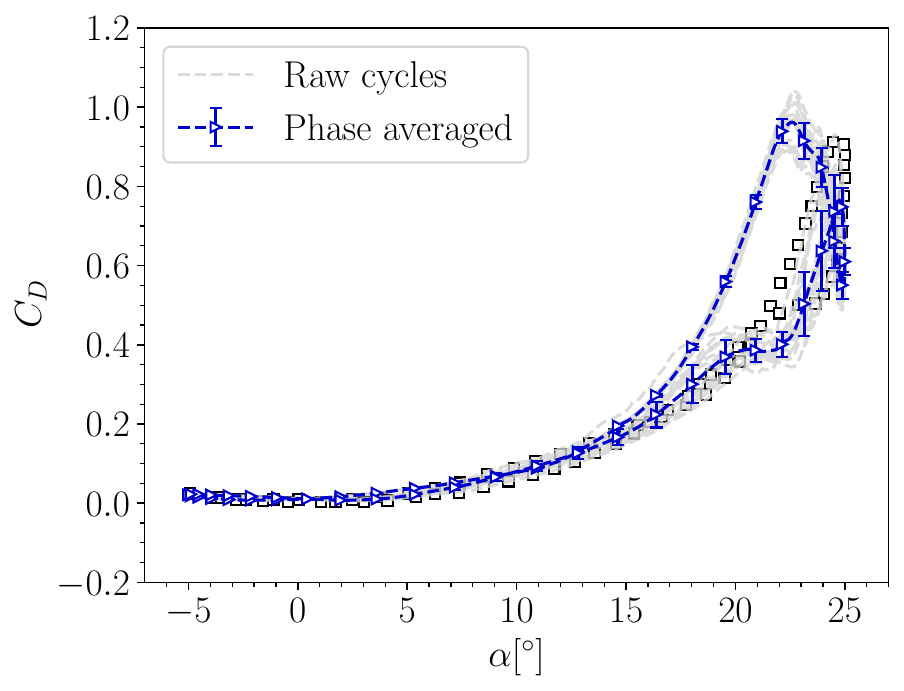}}\\
	
	\caption{Phase-averaged lift and drag coefficients for SBES simulations~\citep{Menter2018}. The average is computed over 16 cycles. The bars indicate the standard deviation. Experimental data from~\citet{Lee2004}.}
	\label{img_sbes_phase_avg}
\end{figure*}

\begin{figure}
	\centering
	\includegraphics[width=0.5\linewidth]{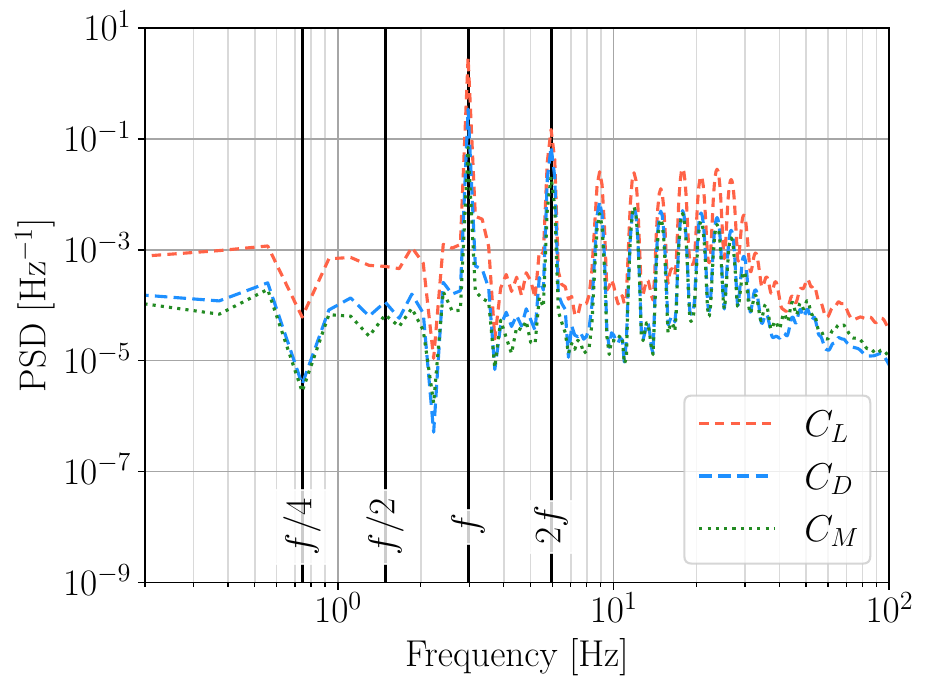}
	
	\caption{Spectral analysis of lift and drag coefficients  for SBES simulations~\citep{Menter2018}. The Power Spectral Density (PSD) is computed over 16 cycles. The obtained spectra are filtered through the algorithm presented in~\citet{Konno1998}}
	\label{img_sbes_periodogram}
\end{figure}

Additional evaluations were conducted for the SBES model.
These evaluations include wall coordinate values at maximum incidence, phase-averaged loads calculated over 16 pitch cycles, and a spectral analysis of force and moment coefficients.
This multifaceted approach provides a deeper understanding of flow dynamics by highlighting the possibilities offered by the turbulence model.

Figure~\ref{img_sbes_plus_unit} depicts the distribution of wall unit values along the profile in the normal $\eta^+$, span-wise $z^+$, and stream-wise $\xi^+$ directions at the maximum incidence angle of 25 degrees. 
All three profiles exhibit a peak value near the leading edge of the airfoil attributable to the high acceleration of the free-stream in this region.
This phenomenon is particularly pronounced in the normal and span-wise directions due to the constant cell size in the first layer.
Nevertheless, the maximum values attained remain within the advised bounds~\citep{Menter2012}.
Additionally, the wall unit values in the span-wise direction are set as half of those observed in the stream-wise direction, along which the flow develops.

An additional advantage of hybrid methods, leveraging the LES portion of the flow solution, is the ability to capture the cycle-to-cycle variability of the loads.
Figure~\ref{img_sbes_phase_avg} presents the trends observed in the lift and drag coefficients over 16 cycles.
The figure additionally includes the phase-average and the corresponding standard deviation.
As anticipated, the pitch-up portion of the cycle typically coincides with attached flow, resulting in load values independent of the pitching cycle.
When the laminar separation bubble detaches, the flow exhibits a more variable pattern, reaching its maximum during the downward pitching phase when the DSV is convected past the body.
Finally, near an angle of 5 degrees, flow reattachment occurs, and the loading trend converges showing the same values.
While SBES model effectively resolves the large-scale turbulent structures for most of the flow and captures the variations between subsequent cycles, it is important to notice that discrepancies with experimental data still exist.

Finally, Figure~\ref{img_sbes_periodogram} depicts the results of a spectral analysis performed on the force and moment coefficients.
The power spectral densities (PSDs) were obtained via a Fast Fourier Transform (FFT) and subsequently filtered using the algorithm proposed by~\citet{Konno1998}. 
In detail, a bandwidth of 100 Hz was adopted to preserve the key spectral peaks in the filtering process.
The analysis reveals a remarkable consistency in the frequency content across all type loads, namely $C_L$, $C_D$, and $C_M$.
As anticipated, the pitching frequency exhibits the highest corresponding PSD.
Additionally, the harmonic frequencies manifest as prominent peaks within the spectrum.
Furthermore, an examination of frequencies lower than the oscillation one reveals the absence of significant peaks.
However, a noteworthy observation is the presence of a PSD value of $10^{-3}$ at half the pitching frequency, which can be attributed to variations associated with fully stalled flow conditions.

\subsection{Comparison with other numerical simulations}\label{ssec_other_simulations}

\begin{figure*}
	\centering
	\subfloat{\includegraphics[width=0.48\textwidth]{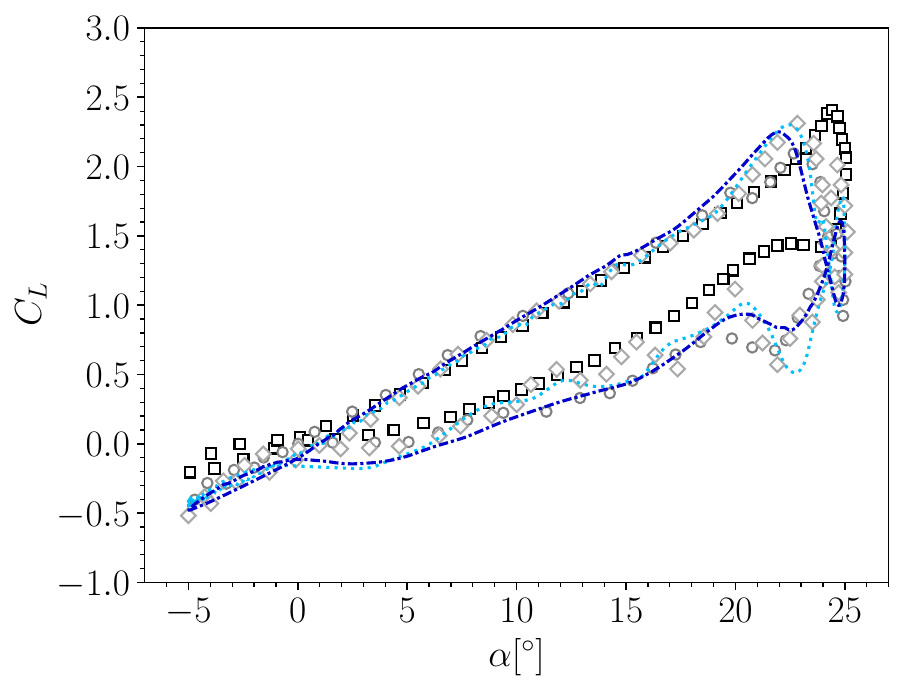}}
	\hfill
	\subfloat{\includegraphics[width=0.48\textwidth]{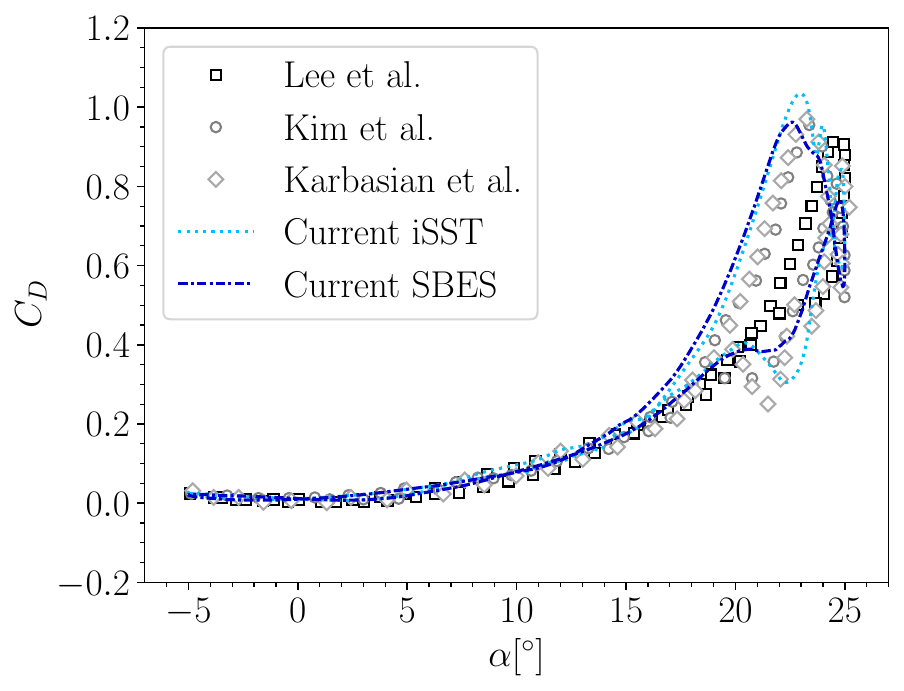}}\\
	
	\caption{Comparison of lift and drag coefficients for current RANS with SST with intermittency turbulence model and hybrid SBES simulations~\citep{Menter2018} with other numerical simulations, \citet{Karbasian2016, Kim2016}, and experimental data, \citet{Lee2004}.}
	\label{img_sim_comp}
\end{figure*}

To conclude the comparative analysis, the turbulence models exhibiting the most promising performance are benchmarked against other numerical simulations documented in the literature.
This comparison encompasses the following: the present SST model with the intermittency equation for the 2D URANS simulation, the SBES model employed in the 3D case, the simulation presented by~\citet{Karbasian2016} utilizing a transition $k-\omega$ SST model, and the LES simulation carried out by \citet{Kim2016}.
It is noteworthy that a high degree of consistency is observed among all the numerical simulations, despite exhibiting the same differences when compared to the experimental data.
Numerical simulations show discrepancies in the angle of DSV generation and the peak values attained by the lift and drag coefficients.
In addition, they present a hysteresis loop near the maximum angle of incidence, which is not present in the experimental data documented by~\citet{Lee2004}.

\section{Conclusions}\label{sec_conclusion}
A computational investigation was undertaken to analyze the flow structures characteristic of dynamic stall.
The primary aim was to conduct a rigorous evaluation of the impact of various turbulence modeling approaches on the behavior of key phenomena, including laminar separation bubbles and dynamic stall vortex generation.
The findings demonstrated that the SST model, coupled with the intermittency equation within the Reynolds-Averaged Navier-Stokes framework, exhibited superior performance in capturing the transitional flow behavior.
The transition plays a crucial role in accurately describing the boundary layer development and the laminar separation bubble.
While dynamic stall is inherently three-dimensional, the study revealed that two-dimensional simulations can effectively represent all essential flow features.
Additionally, the application of state-of-the-at hybrid RANS/LES models was explored.
These models provided different solutions for subsequent pitching cycles enabling studies through numerical simulations at frequency lower than the pitching one.
Despite the advancements in numerical methodologies, discrepancies between the computational results and experimental data remain.
These discrepancies are potentially attributed to the inherent presence of a, albeit small, RANS portion in all simulations.
Within RANS regions, turbulence is entirely modeled, which might not fully capture the complex dynamics of unsteady separation and reattachment observed during dynamic stall.
Supporting this observation, it is evident that numerical simulations employing diverse approaches yield comparable results even when a substantial portion of the flow resolves the largest turbulence structures, as exemplified by the SBES case.
To delve deeper into these discrepancies, an investigation leveraging wall-resolved Large Eddy Simulations should be performed.

\section*{CRediT authorship contribution statement}

\textbf{Giacomo Baldan}: Conceptualization, Methodology, Software, Validation, Formal analysis, Investigation, Data curation, Writing - original draft, Writing – review \& editing, Visualization. \textbf{Francesco Manara}: Software, Validation, Writing – review \& editing. \textbf{Gregorio Frassoldati}: Software, Validation, Writing – review \& editing. \textbf{Alberto Guardone}: Writing – review \& editing, Supervision.

\section*{Declaration of competing interest}
The authors declare that they have no known competing financial interests or personal relationships that could have appeared to influence the work reported in this paper.

\section*{Data availability}
The data that support the findings of this study are available from the corresponding author upon reasonable request.

\section*{Acknowledgments}
The authors acknowledge Leonardo SpA -- Helicopter Division for granting access to the \textit{davinci-1} supercomputer.

\bibliographystyle{unsrtnat}

\end{document}